{\documentstyle[11pt]{article}
\input{sect}
\begin{document}
\title{Analysis of abelian gauge theory with four fermi interaction at
$O(1/N^2)$ in arbitrary dimensions.}
\author{J.A. Gracey, \\ Department of Applied Mathematics and Theoretical
Physics,\\ University of Liverpool,\\ P.O. Box 147,\\ Liverpool,\\ L69 3BX,\\
United Kingdom.}
\date{}
\maketitle
\vspace{5cm}
\noindent
{\bf Abstract.} An arbitrary dimensional expression is given for the anomalous
dimension of the fermion field in a model with a four point interaction and a
$U(1)$ gauge field, at $O(1/N^2)$ within a large flavour expansion in the
Landau gauge.

\vspace{-16cm}
\hspace{10cm}
{\bf {LTH-297}}
\newpage
\sect{Introduction.}
The use of the conventional large $N$ expansion to analyse (renormalizable)
quantum field theories has proved to be successful in revealing properties
which would otherwise be inaccessible in conventional perturbation theory.
The method relies, essentially, on expanding the effective action in the
saddle point approximation, which then determines the leading order structure
in $1/N$. It is virtually impossible, however, to push that analysis beyond the
leading order and probe corrections at $O(1/N^2)$. One method of overcoming
this shortcoming was developed in \cite{1,2} for the $O(N)$ bosonic $\sigma$
model, and applied to other models in \cite{3,4,5}. It is based on solving the
skeleton Dyson equations at the $d$-dimensional critical point of the theory. A
major simplification of this approach is the use of massless fields at this
conformal point of the theory. Hence, one can systematically solve for the
critical exponents, which depend only on $N$ and the space-time dimension, to
several orders beyond the first which represents significant progress. Further,
the absence of a mass allows one to make use of the powerful technique of
uniqueness, introduced in \cite{6}, to evaluate the integrals which arise at
higher orders.

Whilst the initial application of this self consistency method was to a bosonic
model, the techniques have also been developed to probe fermionic theories,
such as the Gross Neveu model, \cite{3}, and quantum electrodynamics, (QED),
\cite{7}, both to $O(1/N^2)$ by computing the critical exponents corresponding
to the anomalous dimension of the basic fermionic field. Other critical
exponents corresponding to the anomalous dimensions of the $3$-vertices, as
well as the $\beta$-function, have also been deduced, \cite{8}. More recently,
we examined a fermionic model involving a four fermi interaction coupled to a
$U(1)$ gauge field at leading order, $O(1/N)$, which is an amalgam of the Gross
Neveu model and QED, \cite{8}. Currently, there is renewed interest in
understanding four fermi interactions, \cite{9}, in addition to earlier work,
\cite{10}, in an attempt to provide insight into the possible composite nature
of the Higgs boson in the standard model, \cite{11}. In this paper, we
therefore extend the $O(1/N)$ analysis of \cite{8} by computing the $O(1/N^2)$
corrections to the critical exponent corresponding to the fermion anomalous
dimension in arbitrary dimensions in the Landau gauge, which will provide a
more realistic scenario for determining the effects the presence of a gauge
field has on the anomalous dimension, rather than studying four fermi
interactions in isolation. Aside from this motivation, such a calculation will
serve as a basis for a future discussion of the Nambu Jona-Lasinio model
coupled to QED, (see, for example, \cite{12}). Further, we will gain insight
into the three dimensional structure of the model, which can be deduced
simultaneously from the present calculation and which has been examined in
\cite{13}.

The paper is structured as follows. In section 2, we introduce our basic
formalism and review the leading order critical point analysis for the model
we are interested in, which we formally extend to $O(1/N^2)$ in section 3. We
discuss the techniques to evaluate the relevant two loop massless Feynman
integrals in section 4 and present our results in section 5.

\sect{Basic formalism.}
We begin by introducing the formalism required to solve the Schwinger Dyson
equation at criticality. First, the (massless) lagrangian we use is, \cite{8},
\begin{equation}
L ~=~ i \bar{\psi}^i \partialslash \psi^i - \frac{(F_{\mu\nu})^2}{4e^2}
+ A_\mu \bar{\psi}^i\gamma^\mu\psi^i + \rho \bar{\psi}^i \psi^i
- \frac{\rho^2}{2g^2}
\end{equation}
where $1$ $\leq$ $i$ $\leq$ $N$, with $N$ playing the role of our expansion
parameter, $A_\mu$ is the $U(1)$ gauge field, $F_{\mu\nu}$ $=$
$\partial_\mu A_\nu$ $-$ $\partial_\nu A_\mu$ and the coupling constants
$e$ and $g$ appear in their respective kinetic terms, in anticipation of the
method we will use, \cite{1,2}. The four fermi interaction can be introduced
explicitly by eliminating the auxiliary field, $\rho$, by its equation of
motion, though we will use the lagrangian, (1), since the method of \cite{1,2}
applies only to theories with $3$-point interactions. As in previous models we
introduce the general structure of the fields at the critical point of the
theory, consistent with Lorentz symmetry, in coordinate space, where the
fields are massless. As our model invloves a $U(1)$ gauge field, we choose to
work in a particular (covariant) gauge, the Landau gauge. This choice is
motivated by the nature of the large $N$ expansion, which is a reordering of
perturbation theory such that chains of bubbles are summed first. Thus one
must be careful to choose a gauge, which is neither affected by this
resummation nor altered by perturbative renormalization effects, \cite{14}.
In the critical region, we therefore take the following asymptotic scaling
forms for the propagators of the three fields of (2.1), as, \cite{1,3,15},
\begin{eqnarray}
\psi(x) & \sim & \frac{A\xslash}{(x^2)^\alpha} ~~~,~~~ \rho(x) \sim
\frac{B}{(x^2)^\beta} \nonumber \\
A_{\mu\nu}(x) & \sim & \frac{C}{(x^2)^\gamma} \left[ \eta_{\mu\nu}
+ \frac{2\gamma}{(2\mu-2\gamma-1)} \frac{x_\mu x_\nu}{x^2} \right]
\end{eqnarray}
as $x$ $\rightarrow$ $0$, where $A$, $B$ and $C$ are the amplitudes of the
respective fields, $\alpha$, $\beta$ and $\gamma$ are their critical exponents
and $d$ $=$ $2\mu$ is the dimension of space-time. Using the Fourier transform
\begin{equation}
\frac{1}{(x^2)^\alpha} ~=~ \frac{a(\alpha)}{\pi^\mu 2^{2\alpha}} \int_k
\frac{e^{ikx}}{(k^2)^{\mu-\alpha}}
\end{equation}
and its derivatives, where $k$ is the conjugate momentum and $a(\alpha)$ $=$
$\Gamma(\mu-\alpha)/\Gamma(\mu)$, it is easy to check that (2.2) indeed
possesses the usual Landau gauge structure, $P_{\mu\nu}(k)$ $=$
$\eta_{\mu\nu}$ $-$ $k_\mu k_\nu/k^2$. From a dimensional analysis of the
action with lagrangian (2.1) we introduce the anomalous pieces of each
exponent by further defining
\begin{equation}
\alpha ~=~ \mu + \half \eta ~~~,~~~ \beta ~=~ 1 - \eta - \chi_\rho ~~~,~~~
\gamma ~=~ 1 - \eta - \chi_A
\end{equation}
where $\eta$ is the (gauge-dependent) fermion anomalous dimension which we
determine at $O(1/N^2)$ here. The remaining relations are deduced from the
dimensions of the $3$-vertices of the action with (2.1) as its lagrangian,
where $\chi_\rho$ and $\chi_A$ are the anomalous dimensions, respectively, of
the interactions involving $\rho$ and $A_\mu$. In previous work, \cite{3}, we
computed each at $O(1/N)$ by carrying out a leading order renormalization
precisely at the critical point using a method based on \cite{16}. This
preliminary analysis was necessary for providing independent checks on the
present calculation. We record that with $\eta$ $=$ $\sum_{i=1}^\infty
\eta_i/N^i$, etc,
\begin{eqnarray}
\eta_1 &=& \frac{(4\mu^2 - 10\mu + 5)\Gamma(2\mu-1)}{4\Gamma^3 (\mu)
\Gamma(2-\mu)} \\
\chi_{\rho \, 1} &=& - \, \frac{(4\mu^2 - 6\mu + 3)}{(4\mu^2 - 10\mu + 5)}
\eta_1 \\
\chi_{A \, 1} &=& - \, \eta_1
\end{eqnarray}
with the convention, $\mbox{tr} 1$ $=$ $4$.

In addition to (2.2) we will require the asymptotic scaling forms of the
$2$-point functions, $\psi^{-1}$, $\rho^{-1}$ and $A_{\mu\nu}^{-1}$. These are
derived from (2.2), by inverting those functions in momentum space before
mapping back to coordinate space. Thus, \cite{1,3,8},
\begin{eqnarray}
\psi^{-1}(x) & \sim & \frac{r(\alpha-1) \xslash}{A (x^2)^{2\mu-\alpha - 1}}
{}~~,~~~ \rho^{-1} (x) ~ \sim ~ \frac{p(\beta)}{B(x^2)^{2\mu-\beta}} \nonumber
\\
A^{-1}_{\mu\nu}(x) & \sim & \frac{m(\gamma)}{C(x^2)^{2\mu-\gamma}} \left[
\eta_{\mu\nu} + \frac{2(2\mu-\gamma)}{(2\gamma-2\mu -1)} \frac{x_\mu
x_\nu}{x^2}
\right]
\end{eqnarray}
where
\begin{equation}
p(\alpha) ~=~ \frac{a(\alpha-\mu)}{\pi^{2\mu} a(\alpha)} ~~,~~
r(\alpha) ~=~ \frac{\alpha p(\alpha)}{(\mu-\alpha)} ~~,~~
m(\gamma) ~=~ \frac{[4(\mu-\gamma)^2 -1] p(\gamma)}{4(\mu-\gamma)^2}
\end{equation}
To obtain the expression for $A^{-1}_{\mu\nu}(x)$ we have first transformed
$A_{\mu\nu}(x)$ to momentum space using (2.3) and inverted it on the transverse
subspace, since this is the physically important part, \cite{16,17}, before
mapping back to coordinate space.

As a preliminary to our $O(1/N^2)$ analysis we illustrate the method by
deriving $\eta_1$ at leading order in coordinate space. First, the skeleton
Dyson equations, with dressed propagators, for each field of (2.1) are
illustrated in figs 1-3. For the moment we will ignore the two loop
corrections and concentrate on the one loop graphs. As each equation is valid
in the critical region, we can represent them by replacing the lines of each
graph by (2.2) and (2.8), since these will dominate. Thus,
\begin{eqnarray}
0 &=& r(\alpha-1) + z + \frac{2(2\mu-1)(\gamma-\mu+1)y}{(2\mu-2\gamma-1)} \\
0 &=& p(\beta) + 4Nz \\
0 &=& m(\gamma) \left[ \eta_{\mu\nu} + \frac{2(2\mu-\gamma)}{(2\gamma-2\mu-1)}
\frac{x_\mu x_\nu}{x^2} \right]
- 4Ny \left[ \eta_{\mu\nu} - \frac{2x_\mu x_\nu}{x^2} \right]
\end{eqnarray}
where $z$ $=$ $A^2B$ and $y$ $=$ $A^2C$. To isolate the physically relevant
part of the photon equation we map (2.12) to momentum space and project out
with the operator $P_{\mu\nu}(k)$ before mapping back again to give,
\begin{equation}
0 ~=~ \frac{(\gamma-\mu)m(\gamma)}{(2\gamma-2\mu-1)}
\end{equation}
As the only unknowns are $\eta$, $y$ and $z$ to leading order, eliminating
the latter two from (2.10), (2.11) and (2.13) yields (2.5), whence
\begin{equation}
y_1 ~=~ - \, \frac{(2\mu-3)\Gamma(2\mu)}{16\Gamma(\mu)\Gamma(2-\mu) \pi^{2\mu}}
{}~~,~~~ z_1 ~=~ - \, \frac{\Gamma(2\mu-1)}{4\Gamma(1-\mu)\Gamma(\mu-1)
\pi^{2\mu}}
\end{equation}
The important part of this exercise aside from illustrating the simplicity of
the technique over the conventional large $N$ renormalization, is to provide
the foundation for proceeding to higher orders. In part this entails expanding
each term of (2.10)-(2.12) to a subsequent order and also including the higher
order graphs of figs. 1-3.

\sect{Corrections to basic formalism.}
The procedure to include the higher order two loop graphs of figs 1-3 is not
straightforward. Whilst it is still valid to represent the lines of the graphs
by (2.8), it follows from their explicit evaluation, which we discuss later,
that the graphs are infinite. These infinities arise from the presence of
divergent vertex subgraphs when the vertex anomalous dimensions are zero. To
handle this we introduce a regulator by shifting the exponents of both the
gauge field and $\rho$ by the infinitesimal quantity $\Delta$, setting
$\beta$ $\rightarrow$ $\beta$ $-$ $\Delta$ and $\gamma$ $\rightarrow$
$\gamma$ $-$ $\Delta$. Consequently, a renormalization procedure is
required which will give a finite set of corrected consistency equations. We
illustrate this in detail for the fermion whose Dyson equation is given
in fig. 1. First, we denote by $\Sigma_i$ the value of the respective higher
order graphs, by which we mean that function of the exponents obtained by
computing the integral with unit amplitudes and without symmetry factors. (We
note that Furry's theorem for (2.1) has excluded various three loop graphs
which are non-zero at $O(1/N^2)$ in other models.) Then we represent fig. 1
near criticality as
\begin{eqnarray}
0 &=& r(\alpha-1) + z u^2 (x^2)^{\chi_\rho + \Delta}
+ y v^2 f(\gamma-\Delta) (x^2)^{\chi_A + \Delta} \nonumber \\
&+& z^2 (x^2)^{2\chi_\rho + 2\Delta} \Sigma_1 + y^2 (x^2)^{2\chi_A + 2\Delta}
\Sigma_2 + 2yz (x^2)^{\chi_\rho + \chi_A + 2\Delta} \Sigma_3
\end{eqnarray}
where $f(\gamma)$ $=$ $2(2\mu-1)(\gamma-\mu+1)/(2\mu-2\gamma-1)$ and we note
that we have not cancelled the powers of $x^2$ since they contribute at
$O(1/N^2)$. We have set $\Sigma_3$ $=$ $\Sigma_4$ which can be observed from
the explicit calculation or by making a change of variables in the integral
itself. In (3.1) we have introduced the vertex counterterms $u$ and $v$
for $\rho \bar{\psi} \psi$ and $A_\mu \bar{\psi} \gamma^\mu \psi$,
respectively, which will be required for removing the infinities. To display
these divergences explicitly we define
\begin{equation}
\Sigma_i ~=~ \frac{K_i}{\Delta} ~+~ \Sigma_i^\prime
\end{equation}
where the prime, ${}^\prime$, denotes the purely finite part of $\Sigma_i$
with respect to $\Delta$. To fix the counterterm we isolate all $1/\Delta$
pieces contributing to (3.1) by expanding each term in powers of $\Delta$ and
setting this to zero, ie
\begin{equation}
0 ~=~ 2 u_1 z_1 + \frac{4(2\mu-1)(2-\mu)v_1 y_1}{(2\mu-3)}
+ \frac{z_1^2K_1}{\Delta} + \frac{y_1^2K_2}{\Delta} + \frac{2y_1z_1K_3}
{\Delta}
\end{equation}
which corresponds to a minimal choice. Absorbing a finite piece will not
affect the value of any exponent we compute. Whilst this choice allows us to
set $\Delta$ $\rightarrow$ $0$ we are still unable to probe the critical
region $x$ $\rightarrow$ $0$ due to the presence of $\ln x^2$ terms which
arise from the non-cancellation of powers of $x^2$ in (3.1). To remove these
singular terms one exploits the freedom in choosing the as yet unspecified
vertex anomalous dimensions. Thus defining
\begin{equation}
z_1 \chi_{\rho \, 1} + f(\gamma) y_1 \chi_{A \, 1} ~=~
- \, [ z_1^2 K_1 + y_1^2 K_2 + 2y_1 z_1 K_3 ]
\end{equation}
we obtain the formal finite representation of the Dyson equation valid in the
critical region as
\begin{equation}
0 ~=~ r(\alpha-1) + z + f(\gamma)y - 2 f^\prime(\gamma)y(v-1)\Delta
+ z^2 \Sigma_1^\prime + y^2 \Sigma_2^\prime + 2yz \Sigma_3^\prime
\end{equation}
One check on this renormalization will come from substituting the explicit
values for the poles into (3.4) and showing it agrees with (2.6) and (2.7).

The analysis for $\rho$ proceeds along analogous lines and its finite
consistency equation is
\begin{equation}
0 ~=~ \frac{p(\beta)}{N} + 4z - z^2 \Gamma_1^\prime - yz \Gamma_2^\prime
\end{equation}
where $\Gamma_i$ $=$ $G_i/\Delta$ $+$ $\Gamma_i^\prime$ and we have defined
\begin{equation}
\chi_{\rho \, 1} ~=~ \quarter [ z_1 g_1 + y_1 G_2 ]
\end{equation}
The minus signs in (3.6) arise from the factor associated with a fermion loop.

The treatment of the $A_\mu$ field, however, requires more care. First, the
Dyson equation of fig. 3 is represented by
\begin{eqnarray}
0 &=& \frac{m(\gamma-\Delta)}{N(x^2)^{2\mu-\gamma+\Delta}} \left[
\eta_{\mu\nu} + \frac{2(2\mu-\gamma+\Delta)}{(2\gamma-2\mu-1-2\Delta)}
\frac{x_\mu x_\nu}{x^2} \right] \\
&-& \frac{4yv^2}{(x^2)^{2\alpha}} \left[ \eta_{\mu\nu}
- \frac{2x_\mu x_\nu}{x^2} \right] - \frac{y^2 \Pi_{1 \, \mu\nu}}
{(x^2)^{4\alpha+\gamma-2\mu-2-\Delta}} - \frac{yz \Pi_{2 \, \mu \nu}}
{(x^2)^{4\alpha+\beta-2\mu-2-\Delta}} \nonumber
\end{eqnarray}
To obtain the correction to (2.13) from (3.8) we first map (3.8) to momentum
space and project out the physically relevant transverse component before
cancelling off a common power of $x^2$, \cite{7}, and then proceed with the
renormalization. Defining $\Pi_{i \, \mu \nu}$ $=$ $\Pi_i \eta_{\mu\nu}$ $+$
$\Xi_i x_\mu x_\nu/x^2$ and the residues at the $\Delta$-poles as $P_i$ and
$X_i$, respectively, we have
\begin{eqnarray}
0 &=& \frac{2(\mu-\gamma)m(\gamma)}{(2\mu-2\gamma+1)N} - \frac{8(\alpha-1)y}
{(2\alpha-1)} - y^2 \left( \Pi^\prime_1 + \frac{\Xi^\prime_1}{2(2\alpha-1)}
+ \frac{X_1}{2(2\alpha-1)^2} \right) \nonumber \\
&-& yz \left( \Pi^\prime_2 +\frac{\Xi^\prime_2}{2(2\alpha-1)}
+ \frac{X_2}{2(2\alpha-1)^2} \right)
\end{eqnarray}
where
\begin{equation}
\chi_{A \, 1} ~=~ - \, \frac{(2\alpha-1)}{8(\alpha-1)} \left[
y_1 \left( P_1 + \frac{X_1}{2(2\alpha-1)} \right) + z_1 \left( P_2
+ \frac{X_2}{2(2\alpha-1)} \right) \right]
\end{equation}
The appearance of the residues, $X_i$, in (3.9) arise from the
$\Delta$-expansion of a function of the exponents which enters after one has
projected out the momentum space component. Of course, the same vertex
counterterms $u$ and $v$ of (3.1) appear in the respective one loop
graphs of figs. 2 and 3. The procedure to complete the calculation is the
same as at leading order. One eliminates $y$ and $z$ from (3.5), (3.6) and
(3.9) to leave one equation with $\eta_2$ as the only unknown and then
substitutes the explicit values for the $2$-loop graphs.

\sect{Computation of $2$-loop graphs.}
Whilst several of the integrals in (3.5), (3.6) and (3.9) have already been
computed in the constituent models of (2.1), we require the values for the
mixed graphs. To compute these we apply the method of uniqueness of \cite{6}
and the method of subtractions of \cite{1,2}. For completeness, we will
illustrate the procedure by computing $\Pi_{2 \, \mu \nu}$ as an example. The
basic uniqueness integration rule we require for a $\sigma \bar{\psi} \psi$
vertex is displayed in fig. 4, where $\alpha_i$ are arbitrary exponents,
satisfying the uniqueness constraint $\sum_{i=1}^3 \alpha_i$ $=$ $2\mu$ $+$ $1$
and $\nu( \alpha_1, \alpha_2, \alpha_3)$ $=$ $\pi^\mu \prod_{i=1}^3
a(\alpha_i)$. To show the necessity of a regulator, if we set $\Delta$ $=$ $0$
in $\Pi_{2 \, \mu\nu}$ and use the rule of fig. 4 to complete the first
integral, we obtain a final value involving the ill-defined quantity $a(\mu)$
which corresponds to the infinity mentioned earlier. However, the regularized
graph cannot be integrated exactly with this rule since the vertex is no longer
unique. Whilst this makes a total evaluation of $\Pi_{2 \, \mu \nu}$ impossible
for all $\Delta$ it is important to realise that we only need the pole and
finite terms with respect to $\Delta$ for $\eta_2$. To extract these we
subtract a quantity, $A(\Delta)$, from $\Pi_{2 \, \mu \nu} (\Delta)$ which has
the same infinity structure but which can be integrated explicitly for non-zero
$\Delta$. Then the difference $\Pi_{2 \, \mu \nu} (\Delta)$ $-$ $A(\Delta)$
will be finite and therefore calculable at $\Delta$ $=$ $0$. The integral which
satisfies these criteria is
\begin{eqnarray}
A(\Delta) &=& \int_y \int_z \rho_\Delta(y-z) \mbox{tr}[\gamma^\mu \psi(-y)
\psi(y-x) \gamma^\nu \psi(x-y) \psi(z)] \nonumber \\
&+& \int_y \int_z \rho_\Delta(y-z) \mbox{tr}[ \gamma^\mu \psi(-z) \psi(y-x)
\gamma^\nu \psi(x-z) \psi(z)]
\end{eqnarray}
where $y$ and $z$ are the locations of the internal vertices of integration.
Each term of (4.1) can be determined by first integrating with respect to
$y$ for non-zero $\Delta$. The final integration is simplified by the general
result
\begin{eqnarray}
\int_y \frac{\mbox{tr} [\gamma^\mu (-\yslash)(\yslash-\xslash) \gamma^\nu
(\xslash-\yslash)\yslash]}{(y^2)^{\alpha_1} ((x-y)^2)^{\alpha_2}}
&=& \frac{4 \nu(\alpha_1-1, \alpha_2-1, 2\mu-\alpha_1-\alpha_2+2)}{(\alpha_1-1)
(\alpha_2-1)} \nonumber \\
&\times& \left[ [(\alpha_1-2)(\alpha_2-2)+\mu-1] \eta^{\mu\nu} \right.
\nonumber \\
&-& \left. 2(\mu-1)(\alpha_1+\alpha_2-\mu-2) \frac{x_\mu x_\nu}{x^2} \right]
\end{eqnarray}
which is valid for all $\alpha_i$. Thus adding the finite piece to
$A(\Delta)$, we find
\begin{equation}
\Pi_{2 \, \mu \nu} ~=~ \frac{8\pi^{2\mu}}{\mu \Gamma^2(\mu)}
\left[ \frac{1}{\Delta} - \frac{1}{(\mu-1)}\right] \left[ \eta_{\mu\nu}
- \frac{2x_\mu x_\nu}{x^2} \right]
\end{equation}
where we have set the exponents to their leading order large $N$ values,
$\alpha$ $=$ $\mu$, $\beta$ $=$ $\gamma$ $=$ $1$. It is worth noting that when
one transforms (4.3) to momentum space in the context of (3.8), the usual
projection operator $P_{\mu\nu}(k)$ emerges, which confirms that our result is
gauge invariant. The procedure to compute the remaining mixed graphs is
similar to that for $\Pi_{2 \, \mu \nu}$ and we merely quote our results for
each to limit the repetitive nature of the discussion. We found
\begin{eqnarray}
\Sigma_3 &=& - \, \frac{4\pi^{2\mu}(2\mu-1)(\mu-1)}{\mu(2\mu-3)\Gamma^2(\mu)}
\left[ \frac{1}{\Delta} + \frac{(\mu^2 - \mu + 2)}{2\mu(\mu-1)^2}
- \frac{2}{(2\mu-3)} \right] \nonumber \\
\Gamma_2 &=& \frac{16\pi^{2\mu}}{(2\mu-3)\Gamma^2} \left[
\frac{(2\mu-1)}{\Delta} - \frac{2(2\mu-1)}{(2\mu-3)} \right. \nonumber \\
&+& \left. \frac{3}{(\mu-1)} - 3(\mu-1) \hat{\Theta}(\mu) \right]
\end{eqnarray}
where $\hat{\Theta}(\mu)$ $=$ $\psi^\prime(\mu-1)$ $-$ $\psi^\prime(1)$ and
$\psi(\mu)$ is the logarithmic derivative of the $\Gamma$-function. For the
integrals involving an internal gauge field we had to first apply an
integration by parts rule which was introduced in \cite{7}. This allows one to
rewrite the integral in terms of simpler integrals where it is easier to take
the trace over eight $\gamma$-matrices, to yield a set of relatively simple
bosonic two loop graphs to which one can apply the subtraction procedure of
\cite{2}.

Finally, for completeness we present the results obtained for the remaining
integrals, noting that the additional factor of $2$ which appears in
$\Gamma_1$ compared with the analogous integral in \cite{3} is due to our
convention that $\mbox{tr}1$ $=$ $4$. Thus,
\begin{eqnarray}
\Sigma_1 &=& - \, \frac{2\pi^{2\mu}}{(\mu-1)\Gamma^2(\mu)\Delta} \left[ 1
- \frac{\Delta}{2(\mu-1)} \right] \nonumber \\
\Sigma_2 &=& \frac{4\pi^{2\mu}(2\mu-1)}{(2\mu-3)^2 \mu \Gamma^2(\mu)}
\left[ \frac{2(2\mu-1)(\mu-2)^2}{\Delta} \right. \nonumber \\
&+& \left. (2\mu-5)\mu + \frac{4(\mu-1)^2 (\mu-2)}{\mu}
- \frac{8(2\mu-1)(\mu-2)^2}{(2\mu-3)} \right] \nonumber \\
\Gamma_1 &=& \frac{8\pi^{2\mu}}{(\mu-1)\Gamma^2(\mu)\Delta} \left[ 1
- \frac{\Delta}{(\mu-1)} \right] \nonumber \\
\Pi_{1 \, \mu \nu} &=& \frac{16\pi^{2\mu}}{(2\mu-3)\Gamma^2(\mu)}
\left[ \eta_{\mu\nu} - \frac{2x_\mu x_\nu}{x^2} \right] \left[
\frac{(2\mu-1)(2-\mu)}{\mu\Delta} \right. \nonumber \\
&+& \left. 3(\mu-1)\hat{\Theta}(\mu) - \frac{3}{(\mu-1)} + \frac{2(2\mu-1)
(\mu-2)}{(2\mu-3)\mu} \right]
\end{eqnarray}
These have been computed earlier in the separate models, \cite{3,7}, using the
technique described at the start of this section.

\sect{Discussion.}
With these explicit values we can now complete our $O(1/N^2)$ analysis. First,
isolating the residues at the $\Delta$-poles we have verified that the
vertex anomalous exponents of (2.6) and (2.7) re-emerge, which is a partial
check on our renormalization and more importantly on the computation of the
$2$-loop graphs. Further, each exponent is consistent with (3.4). It is
worth noting that the relatively simple result, $\chi_{1 \, A}$ $=$ $- \,
\eta_1$ follows directly from the Ward identity for the photon field which is
similar to what occurs in the pure QED version, \cite{7}. More importantly,
we can eliminate $y_2$ and $z_2$ from our finite consistency equations and
after some straightforward algebra we found the following arbitrary dimensional
expression for $\eta_2$,
\begin{eqnarray}
\eta_2 &=& \frac{(2\mu-1) S^2_d}{4\mu(\mu-1)^2} \left[ 4(\mu-1)^2 \hat{\Psi}
+ 16\mu^2(\mu-4) + 11(9\mu-5) + \frac{4}{(\mu-2)} \right. \nonumber \\
&+& \left. \!\! \frac{8}{\mu} + 3\mu(\mu-1)(4\mu^2-10\mu+5)\left( \hat{\Theta}
- \frac{1}{(\mu-1)^2} - \frac{(2\mu-1)}{3\mu(\mu-1)} \right) \right]
\end{eqnarray}
where we have defined $\hat{\Psi}(\mu)$ $=$ $\psi(2\mu-2)$ $+$ $\psi(3-\mu)$
$-$ $\psi(1)$ $-$ $\psi(\mu)$ and $S_d$ $=$ $a(2-\mu)a^2(\mu-1)/\Gamma(\mu)$.

We conclude with several observations. First, the result (5.1) can be evaluated
in three dimensions to gain new information about that model, and we find
\begin{equation}
\eta ~=~ - \, \frac{2}{\pi^2N} ~+~ \frac{4}{9\pi^4N^2} [142 - 9\pi^2]
\end{equation}
With (5.2) we can now examine the question of whether the $O(1/N^2)$ correction
gives a significant numerical contribution to the value of the exponent. The
approximation we have used to solve (2.1) is to assume $N$ is large. However,
in the conventional approach which determines information solely at leading
order it is never clear for which range of values of $N$ one can obtain
reliable accurate information. For instance, the next to leading order
corrections may be of the same magnitude as leading order for a large range of
$N$ and so the approximation would not be a sensible one. By providing an exact
expression at $O(1/N^2)$ one can gain a much better picture of the convergence
of the series in $1/N$ which is one justification for considering the next to
leading order. For instance, from (5.2) it turns out that the lowest value of
$N$ for which there is a sensible correction is $N$ $=$ $3$ which is smaller
than one would naively expect for a large $N$ approximation. In light of these
remarks we can now compare (5.2) with $\eta$ in the Gross Neveu model and QED
for $N$ $=$ $3$ to ascertain which constituent model governs the quantum
properties of the fermion field. So being careful to have the same trace
convention for each case, we find the respective values for $\eta$ are $- \,
0.04$, $0.06$ and $- \, 0.09$. Thus, it would appear that the dominant
contribution in three dimensions derives from the QED sector. Further, it is
important to note that information concerning the three dimensional model,
(2.1), is presently of interest. Recently, several authors investigated the
implications that a $4$-fermi interaction coupled to QED has in relation to
understanding high $T_c$ superconductivity, \cite{18}. Clearly it is necessary
to deduce accurate information for such models and so to achieve this within
the present formalism and to calculate other critical exponents, such as those
relating to the $\beta$-function, to the {\em same} order as (5.1) the
fundamental quantities $y_2$ and $z_2$ need to be known. These can be deduced
from the relevant consistency equations now that $\eta_2$ is available. Indeed
it is worth remarking that one of the features of our formalism is that the
anomalous dimensions of the fundamental fields of a theory have to be computed
first before attempting to calculate others.

Next, in four dimensions if one expands $\eta_2$ in powers of $\epsilon$ $=$
$\mu$ $-$ $2$, then it appears that the leading contribution to the anomalous
dimension of the amalgam model is from the four fermi interaction rather than
the QED sector, since $\eta_2$ $\sim$ $\epsilon$ for (1) and the Gross Neveu
model, whilst $\eta_2$ $\sim$ $\epsilon^2$ for QED, \cite{7}, similar to
leading order. Finally, we remark that it would now be interesting to see if
the equivalence of a Yukawa theory in $4$ $-$ $\epsilon$ dimensions to a four
fermi interaction in $2$ $+$ $\epsilon$, which was established by Wilson in
\cite{10} and others in \cite{9}, is preserved by the addition of the $U(1)$
gauge field. In providing $\eta$ for (1) at $O(1/N^2)$ here, we have introduced
the first step in such an analysis.

\newpage
\noindent
{\Large {\bf Figure Captions.}}
\begin{description}
\item[Fig. 1.] Skeleton Dyson equation with dressed propagators for $\psi$.
\item[Fig. 2.] Skeleton Dyson equation with dressed propagators for $\rho$.
\item[Fig. 3.] Skeleton Dyson equation with dressed propagators for the
gauge field.
\item[Fig. 4.] Integration rule for $\Pi_{2 \, \mu \nu}$.
\end{description}

\newpage
\begin{thebibliography}{99}
\bibitem{1} A.N. Vasil'ev, Yu.M. Pis'mak \& J.R. Honkonen, Theor. Math. Phys.
{\bf 46} (1981), 157.
\bibitem{2} A.N. Vasil'ev, Yu.M. Pis'mak \& J.R. Honkonen, Theor. Math. Phys.
{\bf 47} (1981).
\bibitem{3} J.A. Gracey, Int. J. Mod. Phys. {\bf A6} (1991), 395.
\bibitem{4} J.A. Gracey, Nucl. Phys. {\bf B348} (1991), 737.
\bibitem{5} J.A. Gracey, Nucl. Phys. {\bf B352} (1991), 183.
\bibitem{6} M. d'Eramo, L. Peliti \& G. Parisi, Lett. Nuovo Cim. {\bf 2}
(1971), 878.
\bibitem{7} J.A. Gracey, Mod. Phys. Lett. {\bf A7} (1992), 1945.
\bibitem{8} J.A. Gracey, J. Phys. {\bf A25} (1992), L109.
\bibitem{9} S.J. Hands, A. Koci\'{c} \& J.B. Kogut, Phys. Lett. {\bf 273B}
(1991), 111; J. Zinn-Justin, Nucl. Phys. {\bf B367} (1991), 105.
\bibitem{10} K.G. Wilson, Phys. Rev. {\bf D7} (1973), 2911.
\bibitem{11} W.A. Bardeen, C.T. Hill \& M. Lindner, Phys. Rev. {\bf D41}
(1990), 1467; A. Hasenfratz, P. Hasenfratz, K. Jansen, J. Kuti \& Y. Shen,
Nucl. Phys. {\bf B365} (1991), 79.
\bibitem{12} R.D. Kenway, Rep. Prog. Phys. {\bf 52} (1989), 1475.
\bibitem{13} G.W. Semenoff \& L.C.R. Wijewardhana, Phys. Rev. {\bf D45} (1992),
1342.
\bibitem{14} A. Palanques-Mestre \& P. Pascual, Comm. Math. Phys. {\bf 95}
(1984), 277.
\bibitem{15} A.N. Vasil'ev, Yu.M. Pis'mak \& J.R. Honkonen, Theor. Math.
Phys. {\bf 48} (1981), 750.
\bibitem{16} A.N. Vasil'ev, M.Yu. Nalimov, Theor. Math. Phys. {\bf 55} (1982),
163; {\bf 56} (1983), 15.
\bibitem{17} A.A. Slavnov \& L.D. Faddeev, `Introduction to the Quantum Theory
of Gauge Fields' (Nauka, Moscow, 1978).
\bibitem{18} A. Kovner \& D. Eliezer, Int. J. Mod. Phys. {\bf A7} (1992), 2775.
\end{thebibliography}
\end{document}